# DESIGN STUDY OF COMPACT THOMSON X-RAY SOURCES FOR MATERIAL AND LIFE SCIENCES APPLICATIONS

E G Bessonov[1], M V Gorbunkov[1], P V Kostryukov[2], Yu Ya Maslova[1], V G Tunkin[2], A A Postnov[1], A A Mikhailichenko[3], V I Shvedunov[2], B S Ishkhanov[2] and A V Vinogradov[1]

[1]P.N. Lebedev Physical Institute of the Russian Academy of Sciences, 53, Leninskiy Prospekt, Moscow, 119991, Russia
[2]Moscow State University, Vorobyevy Gory, Moscow, 119992, Russia
[3]Cornell University, Wilson Laboratory, Ithaca, NY 14853, U.S.A.

**E-mails: vinograd@sci.lebedev.ru, gorbunk@sci.lebedev.ru**

**Abstract**

X-ray generators utilizing Thomson scattering fill in the gap that exists between conventional and synchrotron-based X-ray sources. They are expected to be more intensive than X-ray tubes and more compact, accessible and less expensive than synchrotrons. In this work, two operation modes of Thomson X-ray source are documented: quasi CW (QCW) and a pulsed one are considered for material sciences and medical applications being implemented currently at Synchrotron Radiation (SR) facilities. The system contains a ~50 MeV linac and a few picosecond laser with an average power ~few hundred Watts. The Thomson X-ray source is able to deliver up to $5 \cdot 10^{11}$ photons in a millisecond flash and an average flux of $10^{12}$–$10^{13}$ phot/sec. To achieve these parameters with existing optical and accelerator technology, the system must also contain a ring for storage of $e$-bunches for $10^3$–$10^5$ revolutions and an optical circulator for storage of laser pulses for $10^2$ passes. The XAFS spectroscopy, small animal angiography and human noninvasive coronary angiography are considered to be possible applications of the X-ray source.

## 1. Introduction

An infrared photon, if scattered by a relativistic ~50 MeV electron in a head-on collision, changes its energy from $\hbar\omega_L$~1 eV to $\hbar\omega$~20–30 keV depending on the scattering angle. This process of relativistic Thomson scattering can be used for efficient generation of X-rays if a reliable synchronization of dense electron bunches with high power laser pulses in space-time domain is provided. The radiation sources based on this principle can fill the gap between commercial tube-based X-ray sources and conventional synchrotrons. The related technology as well as application field for such laser-electron X-ray generators (LEXG) was studied extensively during the last 10–15 years [1]-[12]. In the next section we will give a brief overview of existing X-ray generators to find out whether there is a niche for LEXG among contemporary and developing X-ray facilities. The temporal structure, X-ray flux and main technical parameters of QCW LEXG are determined in section 3. In section 4 the application of QCW LEXG in XAFS spectroscopy and 3D imaging are discussed. It is shown that 3D imaging technology can benefit considerably if LEXG is used as an X-ray source for micro-tomography. New equipment will combine the high density resolution of medical tomographs with the high spatial resolution of micro-tomography based on micro-focus sources. Section 5 deals with the pulse repetition mode of LEXG which is necessary for the study of live objects. Possible applications of pulsed LEXG are cardiovascular diagnostics and the study of very small vessels and capillaries. These applications pose the strictest requirements on the generated X-ray flux.

## 2. Overview and motivation

X-ray generation is among the key and frontier technologies in such important fields as medical imaging, material studies, drug development and public security. Conventional laboratory sources and user X-ray facilities utilize radiation of electrons in metals or in magnetic fields. Few tens keV X-ray photons are produced in ~100 keV X-ray tubes or in bending magnets and insertion devices installed in few-GeV electron accelerators. Further we will briefly compare some characteristics of tube and accelerator-based X-ray generators.

(a) X-ray tubes are inexpensive, compact and robust devices which convert electric power to the X-ray power with the efficiency 0.1–1%. Their radiation is nearly isotropic but the main disadvantage is that it is not monochromatic and gives poor possibilities for manipulation with the spectrum. However the advanced apparatus for absorption spectroscopy has been developed. Commercially available tube-based XAFS-spectrometer [13] that is capable of analyzing the compounds of 60 elements is presented in Table 1 as example.

**Table 1.** Parameters of XAFS spectrometer.

| # elements | Spectral resolution | Source type | Power | Sizes | Time for 400 points scan |
|---|---|---|---|---|---|
| 60 | $3 \cdot 10^{-4}$ | Rotating anode | 18 kW | 2x3 m$^2$ | 12 hours |

The main problem is that for achieving spectral resolution and elemental sensitivity needed for a broad class of applications, it requires often to enlarge exposure times due to low X-ray flux.

(b) Conventional accelerator based sources – dedicated synchrotrons offer at least, a 4 orders of magnitude higher, average X-ray flux in the beam and allow spectroscopic analysis of diluted samples and investigation of fast processes in living objects. The spectra acquisition requiring hours of exposure time at the X-ray tube-based apparatus and can be taken in minutes at the synchrotrons. Table 2 demonstrates parameters of 2 largest and 2 middle-size synchrotrons.

**Table 2.** Parameters of synchrotrons.

| Facility | ESRF | SPring-8 | Canadian Light Source | Australian Synchrotron Storage Ring |
|---|---|---|---|---|
| Energy GeV | 6.4 | 8 | 2.9 | 3 |
| Circumference, m | 844 | 1436 | 147.4 | 216 |
| Current, mA | 200 | 99.8 | 250 | 200 |
| Cost, $10$^6$ | > 1000 | > 1000 | 194 | 170 |

There are 70 synchrotrons in operation or are under construction in about 30 countries. However, theirs cost, sizes, power supply demands exceed corresponding values of advanced X-ray tubes by a factor of 200–500.

(c) Higher brightness and femtosecond time structure is provided by free electron lasers (FEL). The experiments at the first X-ray FEL FLASH (DESY, see Table 3) designed for the wavelength range of 50–6.5 nm started in 2005 [14]. Construction of the first user FEL facility, XFEL[15] has begun in Hamburg in 2008. It will combine extreme peak and very high average brilliance in hard X-rays. Six more European countries have projects for building FELs in the near future. Compared with conventional synchrotrons, these facilities present the next step in the technology of accelerators as well as theirs sizes and cost (see Table 3).

**Table 3.** Dimensions of the existing installations.

Design Study of Compact Thomson X-ray Sources for Material and Life Sciences Applications

| Facility | FLASH | XFEL |
|---|---|---|
| X-ray range | 50–6.5 nm | 0.1 nm |
| Electron energy | 730 MeV | 17.5 GeV |
| Overall length | 0.33 km | 3 km |
| Bunch charge | 1 nC | 1nC |
| Bunch duration | 400 fs | 88fs |
| # of bunches per beam pulse | 7200 | 3000 |
| Repetition rate | 10 Hz | 10 Hz |
| Emittance | 2 mm mrad | 1.4 mm mrad |
| Beam pulse current | 5 mA or 2.5 kA in a bunch | 5 mA or 10 kA in bunch |
| Cost | | 1081.6 M€ |

(d) Meanwhile the European Union makes efforts to further develop FEL-based research infrastructure [16]. The three-year project has been supported as a Design Study - the European FEL (EUROFEL) and was completed in 2007. EUROFEL is focused on the basic technologies for the next generation FELs: electron injectors, synchronization, high duty-cycle and CW superconducting accelerators, beam dynamics, seeding and harmonic generation [16], [17]. The goal is high-repetition-rate FELs in soft and hard X-rays (see also [18]). The development of FELs is motivated by one of the fundamental and eternal scientific challenges: to understand and control the structure and properties of materials and living matter.

At the same time, much of fundamental and applied research is routinely made now not only in large facilities, but also at universities, academic and industrial labs, medical schools etc. However, a mid-size university presently cannot afford even a conventional synchrotron. The problem is recognized by the European Commission that approved the COST Action MP0601 "Short Wavelength Laboratory Sources" [19], claiming "…Although expansion in application areas is due largely to modern synchrotron sources, many applications will not become widespread, and therefore routinely available as analytical tools, if they are confined to synchrotrons. This is because synchrotrons require enormous capital and infrastructure costs and are often, of necessity, national or international facilities. This seriously limits their scope for applications in research and analysis, in both academia and industry. How many universities, research institutes or even industrial laboratories would have electron microscopes if electron sources cost 100 M€ or more? Hence there is a need to develop bright but small and (relatively) cheap x-ray sources, not to replace synchrotrons but to complement them."

LEXG based on relativistic Thomson scattering has the potential to become a new X-ray source and fill the gap between X-ray tubes and synchrotrons for a wide variety of fundamental research and practical applications. In the following sections we will consider the LEXG schemes meeting the requirements of material composition analysis, high density resolution micro-tomography, small animal angiography and human noninvasive coronary angiography.

**3. Quasi CW mode**

Let us consider a simplified generator scheme as it is shown in Figure 1. A laser and a source of relativistic electrons (linac) synchronically direct laser pulses and $e$-bunches of approximately equal durations 2–10 ps to an interaction point (IP). To enhance the number of collisions, the laser pulses are trapped for $n_C \approx 100$ revolutions in an optical circulator and electron bunches are stored in a ring of the same perimeter ≈ 3 m for $n_S = 10^3$–$10^5$ revolutions. The values $n_c$ and $n_s$ are limited by radiation damage of optical materials and the lifetime of the $e$-bunch emittance in a storage ring. Thus linac injects a bunch into a ring with the rate $v_e$ and laser with the same rate $v_L = v_e$ emits a train of $n_L = n_s/n_c$ pulses into a circulator. Each head-on collision of a laser pulse (photon energy $\hbar\omega_L = 1.16$ eV) and $e$-bunch (electron energy $E_e = 43$ MeV) results in generation of an X-ray burst with a maximum photon energy $\hbar\omega \approx 33$ keV. A number of X–ray photons generated in one collision is

Design Study of Compact Thomson X-ray Sources for Material and Life Sciences Applications

$$N = N_e N_L \frac{\sigma_T}{s_e + s_L}, \qquad (1)$$

where $N_e$, $N_L$ are the total numbers of electrons and photons respectively, $\sigma_T = 6.6 \cdot 10^{-25}$ cm$^2$, $s_e = 2\pi\sigma_e^2$, $s_L = 2\pi\sigma_L^2$; $\sigma_e$, $\sigma_L$ are transverse sizes of electron and laser beams respectively, $\sigma_e = \sqrt{\varepsilon\beta}$, where $\varepsilon$ stands for emittance and $\beta$ stands for beta-function of electron beam at the Interaction Point (IP). The substantial part of photons (1) is contained in a narrow angle $\sim \gamma^{-1} \approx 0.01$, where $\gamma = E_e/mc^2$.

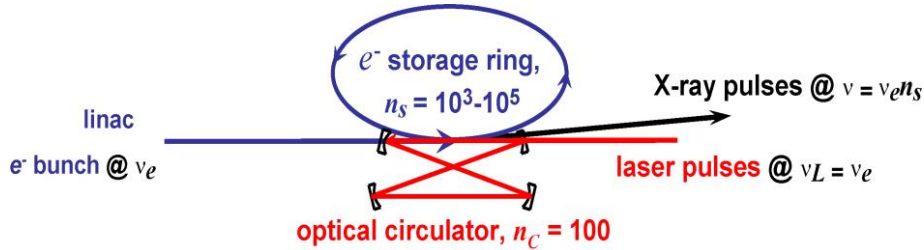

**Figure 1**. A principal scheme for providing efficient $e$-bunch – laser pulse collisions.

For further estimations we will take the following parameters of low emittance Linacs, existing and projected: $N_e = 6 \cdot 10^9$ (1 nC bunch charge) $\gamma\varepsilon_n = 1$ mm·mrad – normalized emittance, $\sigma_e = \sigma_L = 20$ μm. Then by the reasons of avoiding damage of circulator mirrors, reliability and commercial availability of the laser, we restrict the laser micro-pulse energy by 10 mJ and the laser average power $P_L \leq 0.5$ kW. Anyway the X-ray beam consists of picosecond bursts radiated in the collisions of 1 nC bunches with 10 mJ (or less) laser pulses. The total amount of X-ray photons emitted in one "building brick" collision for infra-red laser ($\hbar\omega_L = 1.16$ eV) is

$$N = 1.3 \times 10^{-20} N_L N_e = 7.8 \times 10^{-11} N_L = 5 \times 10^6. \qquad (2)$$

The goal of the design is to arrange these standard collisions in a compact, cost effective system so as to meet the requirements of specific applications.

The X-ray flux $\Phi$ provided by the generator from Figure 1 equals to

$$\Phi = N\nu_e n_S = N\nu_e n_L n_C, \quad P_L = E_L \nu_e n_L = E_L \nu_e \frac{n_S}{n_C}. \qquad (3)$$

The 3 m optical circulator (and the ring) perimeter allows the product $\nu_e n_S$ up to 100 MHz. This is the value of linac frequency considered in the design of the next generation CW light sources involving warm and cryogenic accelerator structures [16, 18]. To achieve it, the problems connected with radiation and electrical damage of cathodes and availability of photocathode UV lasers (see [11]) are to be solved. The storage ring in LEXG reduces the linac frequency down to 0.5–5 kHz for single bunch injection (see further this section) and no problems arise with the cathode and photocathode laser.

To find the design features resulting from limitations on the laser power $P_L = 0.5$ kW and the laser micro-pulse energy $E_L \leq 10$ mJ let us consider two relations following from (2) and (3):

$$\Phi = n_C N_e \frac{\sigma_T}{s_e + s_L} \frac{P_L}{\hbar\omega_L} = 2.5 \times 10^{13} \frac{phot}{\sec}, \quad \nu_e n_S = \frac{P_L}{E_L} n_C \geq 5 \text{ MHz.} \quad (4)$$

Firstly, eq. (4) gives maximum for X-ray photon flux in a QCW mode which is proportional to average laser power. Secondly, it determines the lower limit of the product $\nu_e n_s$. On the other hand $\nu_e$ and $n_s$ by technical reasons cannot be too large to avoid overloading of the photocathode and to achieve the system compactness and reasonable cost. For $E_L = 10$ mJ the realistic compromise is:

$$\nu_e = 500 \text{ Hz}, \quad n_S = 10^4. \qquad (5)$$

Design Study of Compact Thomson X-ray Sources for Material and Life Sciences Applications

Note that for the fixed number of laser pulses per second $v_L n_L = v_e n_s / n_c$ some freedom remains for the time structure of the laser beam, i.e. for the choice of $v_L$ and $n_L$. Various approaches to the laser design for production of trained laser beams for circulators of laser-electron X-ray generators are considered in [6, 11].

So formulas (4) and (5) and other relations represented in this section determine the output flux and main parameters of QCW LEXG. In the next section we will briefly discuss possible applications.

**4. Applications of QCW LEXG**
As was shown in the previous section the LEXG is expected to provide a photon flux $\sim 2 \cdot 10^{13}$ sec$^{-1}$ that is sufficient for applications in material sciences. For example, a standard method to determine material chemical composition and structure is X-ray Absorption Fine Structure Spectroscopy (XAFS) [20]. One XAFS spectrum typically contains ~600 measurements in narrow spectral intervals $\delta\lambda/\lambda \approx 10^{-3}$–$3 \cdot 10^{-4}$ and for practical reasons must take not more than 1–2 hours. This yields that the photon flux before monochromator can be estimated as $10^{12}$ sec$^{-1}$ [20], [21] that is well below the value given by (4).

Another possible field for application of LEXG is X-ray Computer Tomography (CT) which is widely used now for 3D imaging in medicine, industry, applied research etc. Normally for 3D image reconstruction, 100–200 projections are required. Medical CTs provide very high accuracy, ~few tenths of 1 percent of the contrast measurement that is of principal importance for diagnostics of many human diseases. The spatial resolution provided by medical CT reaches ~100 μm whereas standard value is four to five times larger. In practice, spatial resolution is a compromise of several factors: source and pixel sizes, source power, exposure time, absorbed radiation dose etc. However, some kinds of applications require higher spatial resolution than that provided by medical CTs (see the next section). The problem is overcome by *micro-tomographs* (micro CTs), that are based on the combination of table-top micro-focus tubes and CCD detectors [22]. Micro CT is a compact device designed for imaging of ~1 mm–5 cm objects. Due to the reduced source size, its spatial resolution reaches ~ few microns. In fact it is reached at the expense of accuracy of contrast measurements that is limited by relatively low average photon flux produced by micro-focus sources – air-cooled < 50W power X-ray tubes with a 1–10 μm focal spot. So the density resolution of micro CT is now by far lower than that of medical CTs. Therefore it was always a challenge for micro CT to visualize the most essential low contrast tissues for biomedical research (vessels, membranes, blood, nerves, muscles, brains etc.). This can be achieved by illuminating every voxel with a sufficient enough amount of X-ray photons to obtain high signal-to-noise ration and improve density resolution. Hence the X-ray tube has to be replaced by another also micro-focus but more powerful X-ray source.

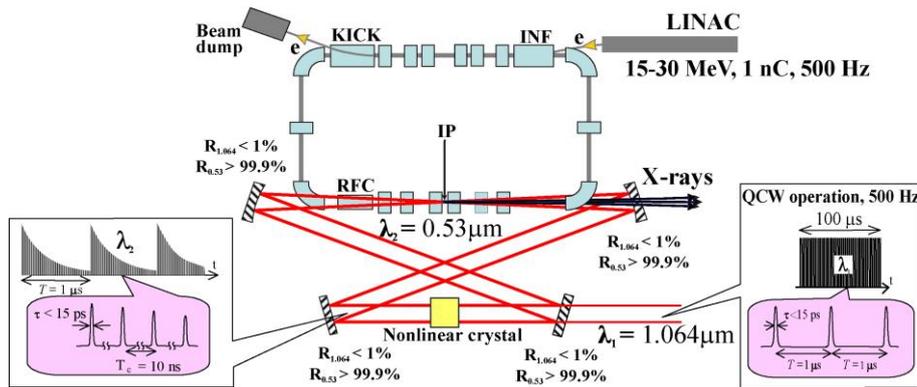

**Figure 2**. QCW Laser-electron X-ray Generator for micro-CT or XAFS applications: circulator based on second harmonic generation in a high-Q cavity.

Taking this into account, the requirements for the new source are the following:
To keep the acquisition time of one projection in the range of 0.1–10 sec (as in standard micro CT) and to increase the accuracy of the absorption measurements the source must provide 100–1000 times more photons delivered and registered in every single pixel of the detector. The number of photons incident on the object and needed for 0.1% accuracy of contrast measurement can be estimated as $10^{12}$ per one projection ($10^5$ pixels).

Design Study of Compact Thomson X-ray Sources for Material and Life Sciences Applications

The flux needed to collect one tomographic dataset (200 frames) in one hour is $\sim 2\cdot10^{11}$ sec$^{-1}$ which is also in reach of LEXG as estimated in the previous section.

Figure 2 shows the principle scheme of such X-ray generator with an optical circulator based on frequency doubling of a pumping IR laser radiation (see [11]).

**5. Pulsed operation mode for life sciences. Coronary angiography**
For some medical and industrial research and applications, the pulsed X-ray beam is more advantageous. Average X-ray power is determined by the same equation (4), but the linac rate $v_e$ depends on the specific application. For example to observe breath, heartbeat and other physiological motion, the exposure 0.1–1 ms and the rate of frames $v = 30$ Hz are needed. In LEXG this is achieved by collision of pulse trains. The resulting 30 Hz X-ray signal consists of trains of $n_S$ micropulses separated by 10 ns. If consider life sciences then $v_e = 30$ Hz that limits laser power (see (3)) and hence the X-ray output:

$$P_L = E_L \frac{v_e n_S}{n_C} \approx 300 \text{ W}, \quad \Phi = v_e N_e \frac{\sigma_T}{s_e + s_L} \frac{P_L}{\hbar \omega_L} = 1.5 \times 10^{13} \frac{phot}{\sec}. \quad (6)$$

As optimistic values $E_L = 10$ mJ and $n_S = 10^5$ have been used in estimate (6), it gives the maximum X-ray flux that can be reached in the pulsed repetition mode. Further we will discuss two possible applications of such a source in angiography that are currently made at synchrotron beam lines.

A number of new medical technologies for imaging and therapy for which the X-ray flux of conventional sources is not sufficient are being developed during 1–2 decades on synchrotron radiation facilities [23, 24]. Partly with the hope that if successful they may be implemented on a new intense, compact and available source appropriate for application in clinics. This could be an important and promising field for expansion of LEXG.

One of these new technologies is aimed at selective coronary angiography which is the most widespread tool for the diagnosis of cardiac diseases. In this invasive method, X-ray images of arteries that are expected for affection or pathology are produced at arteries catheterization for delivering a contrast agent directly to the investigated area. Millions of patients a year undergo this investigation. The procedure is not absolutely safe for patients due to (small) probability of artery damage and sometimes negative side effects caused by contrast agent, and for medical personnel acquiring radiation doses. So the screening is not always recommended even though the diagnostics are very desirable at the earliest stages of ischemic disease. Therefore several non-invasive methods of coronary arteries imaging are being developed [25]. However, currently selective coronary angiography remains the "gold standard" for diagnosis of most heart diseases as other methods do not provide comparable image quality.

A new noninvasive method utilizing synchrotron radiation was suggested to replace selective coronary angiography. It was suggested that the high power of a synchrotron beam allows for the improvement in image quality by digitally subtracting images obtained on two sides of absorption K-edge of iodine contrast agent [26]. This substantially reduces the concentration of the contrast agent and gives possibility to inject it intravenously (noninvasively).  After ~400 tests at synchrotron DORIS, this showed very good image quality and acceptance of the new noninvasive procedure by patients [27]. Then a project of 1.6 GeV storage ring was published as a step towards facilities for noninvasive coronary angiography to be installed in medical centers and clinics [28]. However further progress in this field slowed down.  Probably the reason is that everyday medical exploitation of a facility with GeV-scale storage ring and 42.2x37.6 m sizes is not economically justified.

Design Study of Compact Thomson X-ray Sources for Material and Life Sciences Applications

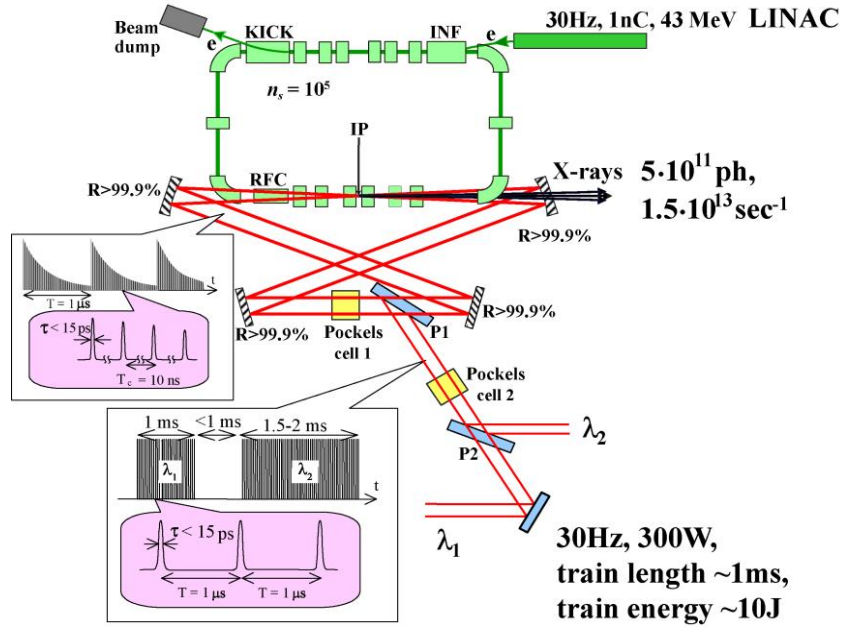

**Figure 3.** A scheme of LEXG designed for human K-edge subtraction imaging angiography.

A conceptual design of LEXG for noninvasive coronary angiography based on subtraction imaging was considered in [29]. The design contained a ~50 MeV linac, 1 m radius storage ring and a picosecond laser. The storage ring increases average current and makes easier the requirements to photo-injector. The usage of trained (~$10^2$–$10^4$ micropulses) laser beams for Thomson scattering allows to decrease X-ray photon flux employed for video screening in human angiography (~$10^{13}$) and also to reduce the radiation dose on a patient. Though the obtained value is in agreement with estimations of other authors [3], it is difficult to say now without an experiment whether this flux will be sufficient for practical angiography. An updated LEXG scheme for noninvasive coronary angiography is given in Figure 3. Two lasers are needed for taking images on both sides of iodine K-edge.

Synchrotron radiation due to high X-ray photon flux is being applied to animal studies for new treatments in medical practice. It enables an *ex vivo* and *in vivo* research with high spatial and temporal resolution. Special synchrotron radiation micro-angiography (SRM) systems have been developed for visualization of vessels and cardiac pathologies in small animals: mice, rats, dogs and rabbits [30]-[32]. The latter, at SPring-8, has the following parameters: field of view 7x7 mm or 4.5x4.5 mm; pixel size 7x7 µm or 4.5x4.5 µm; spatial resolution 6 µm; minimum detectable vessel diameter 20 µm; shortest exposure time 2 ms; rate 30 frames/sec. Note that conventional angiography does not allow visualization vessels with a diameter less than 200 µm. This SRM imaging system was used to evaluate morphometric and physiological evaluation of coronary collateral microvessels development in stem cells transplantation in rats. The 6 µm resolution images were first obtained on beating hearts that was not previously available for conventional angiography as well as for micro CT. For example, coronary collateral growth was studied earlier by micro CT, but with spatial resolution three times larger – 18 µm, and as was discussed in section 4 this could be performed only at postmortem examination [33] with long enough exposures. Now efforts are being done to improve the system in two respects: (a) to reduce the exposure time down to 0.1 ms in order to obtain blur-free images and completely use the resolving potential of the detector (b) to switch from ex-vivo to in-vivo experiments with individual animals. For this purpose the photon flux will be increased with X-ray optical system.

Let us discuss the design of LEXG system for micro-angiography which meets the above mentioned requirements. As SRM utilizes radiation above the iodine K-edge and does not use the subtraction of images, only one laser of two shown in Figure 3 is needed. Then the estimation of the number of photons to obtain one frame gives ~$10^{10}$ which leads to the average photon flux ~$3 \cdot 10^{11}$ sec$^{-1}$. Again we could see that this type of research remains within the reach of LEXG. Special measures can be taken for providing 0.1 ms exposure and for reduction of the source size below 10 micrometers.

Design Study of Compact Thomson X-ray Sources for Material and Life Sciences Applications

## 6. Storage ring and injector

Storage ring is one of the key components of the LEXG that provides the increase in the output flux by a factor of $10^3 - 10^5$, while makes the system substantially complicated and bulky. To avoid these problems we included into the design superconducting (SC) elements. This resulted in the possibility to mount the storage ring and injector on a plate of 1.35m× 1.63m (see Fig.4). The laser system is assembled on a separate optical table (plate) located under the accelerator table or above it. These installations could be rotated together in 3D in horizontal and vertical directions around IP. Slope angle around horizontal axis (providing vertical scan) defined by the allowable slope of SRF helium case which is ~ $\pm 10^o$. Rotation in horizontal (frontal) plane around vertical axis running through IP could be made within $\pm 90^o$. This allows easy scanning of outcoming X ray.

*6.1. Injector*

We suggest usage of superconducting RF (SRF) photo gun and SC accelerating structure operating at ~10 cm wavelength. This eliminates heavy and bulky klystron system for feeding accelerating structure. Complications associated with usage of helium temperature systems pays for compact size, light weight and mobility of the entire system.

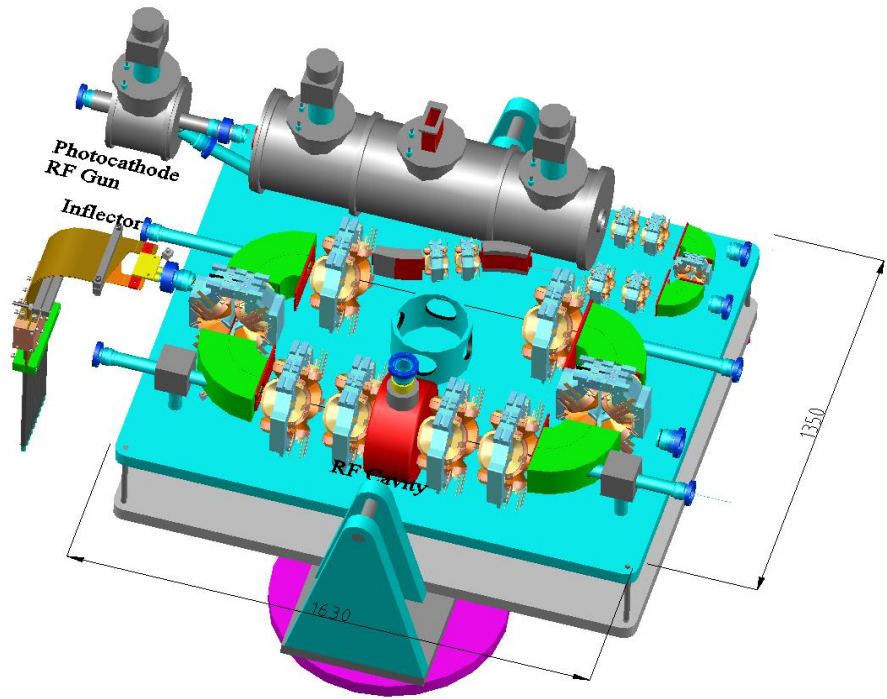

Design Study of Compact Thomson X-ray Sources for Material and Life Sciences Applications

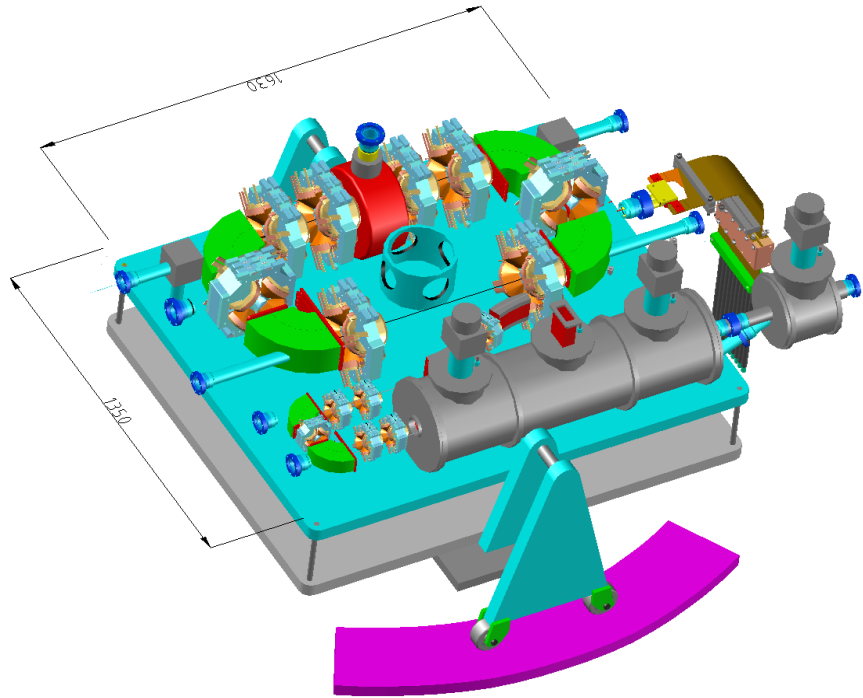

**Figure 4.** Detailed view of Thomson X-ray generator. Laser System mounted on a separate bench mounted under the accelerator plate. Vacuum chamber of storage ring is not shown.

RF Photo gun is suggested as a primary source of electrons. This technique is well developed [35]. Laser system must generate a laser pulse each 1/30 sec (30Hz) synchronized with RF and main laser system.

Length of accelerated structure is ~1 m, wavelength is ~10 cm. Linac is suggested to be a traveling wave multi-cell structure circled from out to the input cavity by SC waveguide. This first, reduces ~two times the ratio of surface to acting RF field strength and second, allows usage of RF structure with broad variety of phase shift per cell and high group velocity. RF power is injected in the circled loop with the help of directional coupler directly in a waveguide running outside of accelerating structure. As the beam load is negligible here, so the external quality factor of ~$10^8$-$10^9$ could be expected here, delivering shunt impedance $Z_\omega \cong 10^{12} Ohm/m$. This allows powering the RF structure for net accelerating voltage $U$=40 MV with $P \cong U^2/Z_\omega \cong$ 1.6 kW of power running inside the loop. This could be driven by vacuum tube of <1 kW.

Cooling by liquid He with two cryo-coolers is a feasible option here, allowing compact design of refrigeration system.

*6.2. Storage ring*

Ring has four magnets with bending radius ~15 cm. Beam optics has a mirror symmetry. Collisions of electron and photon beams occur inside RF cavity, shown in Fig.4. Single-turn radial injection is made with the help of pulsed inflector also shown in Fig. 4 at the left.

The main *e*-bunch parameters are close to those used in Sections 3–5: charge is 1 nC, diameter $\sigma_e$ = 14 μm, emittance $\gamma\varepsilon_{x,y}$ = 1 mm·mrad, $\beta_{x,y}$ = 2 cm, emittance lifetime > 1 ms, bunch length ~2–10 ps.

Design Study of Compact Thomson X-ray Sources for Material and Life Sciences Applications

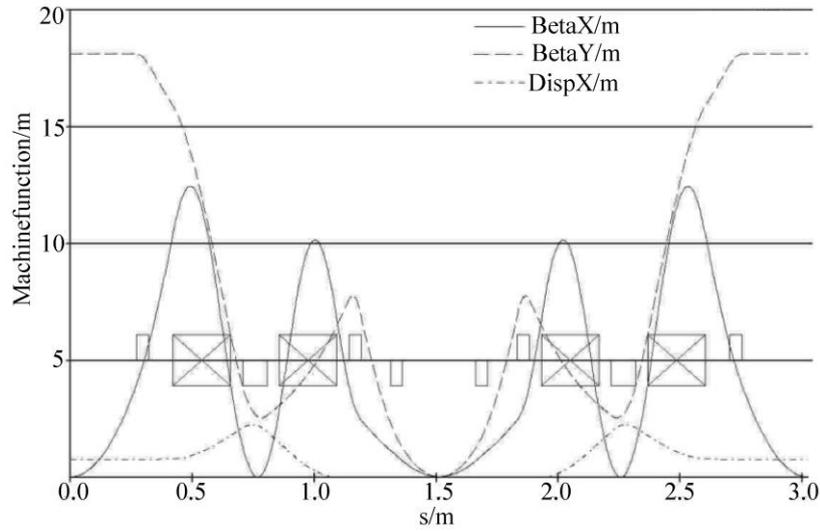

**Figure 5.** Envelope functions around the ring.

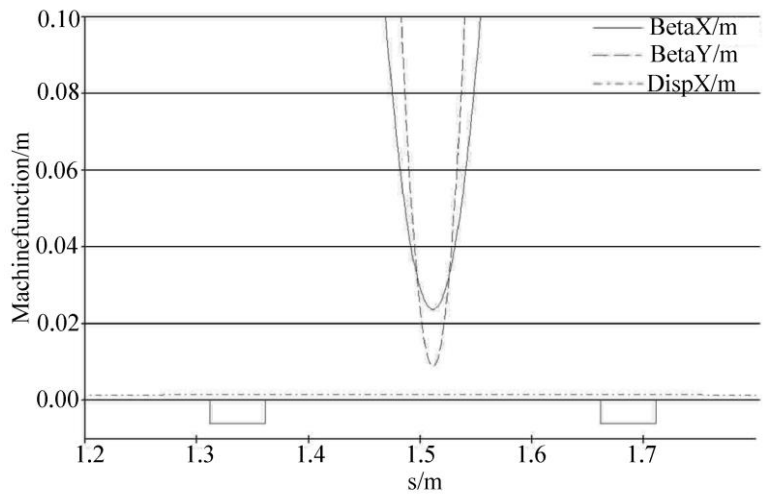

**Figure 6.** Magnified envelope functions around IP.

**Table 4.** Main parameters of Storage Ring.

| | |
|---|---|
| Energy | 30-50 MeV |
| Circumference | 3 m |
| Magnet bending radius | 0.15 m |
| Bunch charge | 1 nC |
| Invariant Emittance | $\gamma\varepsilon_x = 1.1$ mm·mrad |
| in the ring | $\gamma\varepsilon_y = 0.6$ mm·mrad |
| Beam sizes at IP | $\sigma_x = 15\mu m$, $\sigma_y = 8\mu m$, $\sigma_z \approx 3$ mm |
| Toushek lifetime | 1290 sec |

## 7. Discussion and summary

Main characteristics of the two types of Thomson X-ray sources (LEXG) designed for material science applications (quasi CW) and imaging of live objects (pulsed repetitive) are presented in Table 5. The first four rows characterize the output beam of 33 keV photons. The last three concern the IR pumping laser.

Design Study of Compact Thomson X-ray Sources for Material and Life Sciences Applications

**Table 5.** Main parameters of the source suggested.

| Operation mode | quasi CW | Pulsed repetitive |
| --- | --- | --- |
| Average flux, sec$^{-1}$ | $2.5 \cdot 10^{13}$ | $1.5 \cdot 10^{13}$ |
| Rate of flashes = linac operation rate, Hz | 500 | 30 |
| X-ray pulses in a flash | $10^4$ | $10^5$ |
| # of X-ray photons in a flash | $5 \cdot 10^{10}$ | $5 \cdot 10^{11}$ |
| Laser pulse energy, mJ | 10 | 10 |
| Laser average power, W | 500 | 300 |
| Laser pulses in a train | $10^2$ | $10^3$ |

The schemes of X-ray sources considered in this paper contain an optical circulator and a compact electron storage ring. Both are necessary for making LEXG competitive with conventional tube-based sources. The optical circulator with $n_C = 10^2$ revolutions was described in the previous publication [11]. Figure 4 displays the scheme of a compact storage ring.

In conclusion, the designs of two types of laser-electron X-ray generators based on Thomson scattering are presented for control of materials and imaging of live objects. To be realistic, the pulse energy of IR laser was restricted by $\leq 10$ mJ and average power $\leq 0.5$ kW. Other requirements of this study to lasers, accelerators and their synchronization are also close to current technology and well correspond to the level stated in [17]. The realization of compact Thomson scattering X-ray source will make many modern methods developed on synchrotron radiation facilities available to a wide community of material and medical scientists. This perspective and technical feasibility attract more and more researches to the field [34].

**Acknowledgments**
The work was supported by the Program "Laser systems" of Russian Academy of Sciences and RFBR Grant 08-08-00108-a.

Design Study of Compact Thomson X-ray Sources for Material and Life Sciences Applications